%% file: paper.tex
\documentclass{article}
\usepackage{frascatiphys}
\usepackage{epsfig}
\RequirePackage{xspace}
\usepackage{relsize}

\def\BaBar   {\mbox{\slshape B\kern-0.1em{\scriptsize A}\kern-0.1em
              B\kern-0.1em{\scriptsize A\kern-0.2em R}}\xspace}
\def\refBaBar{\mbox{\slshape B\kern-0.1em{\scriptsize A}\kern-0.1em
              B\kern-0.1em{\scriptsize A\kern-0.2em R}}\xspace}
\def\etal    {{\it et~al.}\null\xspace}

\def\epem    {\ensuremath{e^+e^-}\xspace}
\def\g       {\ensuremath{\gamma}\xspace}
\def\W       {\ensuremath{W}\xspace}
\def\qqbar   {\ensuremath{q\overline q}\xspace}
\def\d       {\ensuremath{d}\xspace}
\def\s       {\ensuremath{s}\xspace}
\def\b       {\ensuremath{b}\xspace}
\def\kaon    {\ensuremath{K}\xspace}
\def\Kstar   {\ensuremath{K^*}\xspace}
\def\Dstarz  {\ensuremath{D^{*0}}\xspace}
\def\B       {\ensuremath{B}\xspace}
\def\Bbar    {\kern 0.18em\overline{\kern -0.18em B}{}\xspace}
\def\BB      {\ensuremath{B\Bbar}\xspace} 
\def\Bz      {\ensuremath{B^0}\xspace}
\def\Bzb     {\ensuremath{\Bbar^0}\xspace}
\def\X       {\ensuremath{X}\xspace}
\def\mes     {\mbox{$m_{\rm ES}$}\xspace}
\def\DeltaE  {\mbox{$\Delta E$}\xspace}

\mathchardef\Upsilon="7107
\def\Y#1S{\ensuremath{\Upsilon{(#1S)}}\xspace}% no space before {...}!
\def\FourS {\Y4S}

\newcommand{\gev}{\ensuremath{\mathrm{\,Ge\kern -0.1em V}}\xspace}
\newcommand{\gevcc}{\ensuremath{{\mathrm{\,Ge\kern -0.1em V\!/}c^2}}\xspace}

\newcommand{\jprlBase}       {Phys.\ Rev.\ Lett.\xspace}
\newcommand{\jprBase}        {Phys.\ Rev.\xspace}
\newcommand{\jprl}      [1]  {\jprlBase\ {\bf #1}}
\newcommand{\jprd}      [1]  {\jprBase\ D~{\bf #1}}
\begin{document}

\input slacpub-cover.tex
\title{Rare Decays and Search for New Physics with \BaBar}
\author{
Johannes M. Bauer \\
{\em University of Mississippi, University, MS 38677, U.S.A.} \\
for the \BaBar Collaboration}
\maketitle
\baselineskip=11.6pt
\begin{abstract}
    Rare \B decays permit stringent tests of the Standard Model and
    allow searches for new physics.  Several rare radiative-decay
    studies of the \B meson from the \BaBar collaboration are described.
    So~far no~sign for new physics was discovered.
\end{abstract}
\baselineskip=14pt

\section{Introduction}

At the SLAC PEP-II \B-Factory, the \BaBar detector collected so far more
than $250$M \BB pairs, created by \epem collisions at the \FourS
resonance.  This data set makes searches for rare decays feasible at
branching fractions (BF) of $10^{-4}$ or less.  This talk concentrates
on radiative \B decays.  Additional results from \BaBar were discussed
elsewhere at this conference.\cite{NandoGagan}

\section{Fully- and Semi-inclusive $\B\to X_s\g$, $\B\to\Kstar(892)\g$ 
\& $\B\to\Kstar_2(1430)\g$}

The~lowest-order Feynman diagram of $b\to s\g$ is a one-loop
electromagnetic penguin, in which non-Standard Model (non-SM) virtual
particles (like the Higgs) might influence the decay rate.  Measuring
the energy distribution of the \b quark inside the \B meson helps extract
$|V_{ub}|$ from $\B\to X_ul\nu$.  The decay $\b\to\s\g$ was studied in
inclusive and exclusive modes using $\sim89$M \BB pairs.

In the so-called ``fully-inclusive'' measurement only the photon of
$\B\to\X_s\g$ needs to be detected, but large background has to be
suppressed.  In the ``semi-inclusive'' measurement, the $\B\to\X_s\g$ BF
is determined from 38~exclusive states with about 45\% of the total rate
estimated to be missing.

\begin{figure}[htbp]\vspace{-2mm}
\centerline{\epsfysize 1.45truein
            \epsfbox{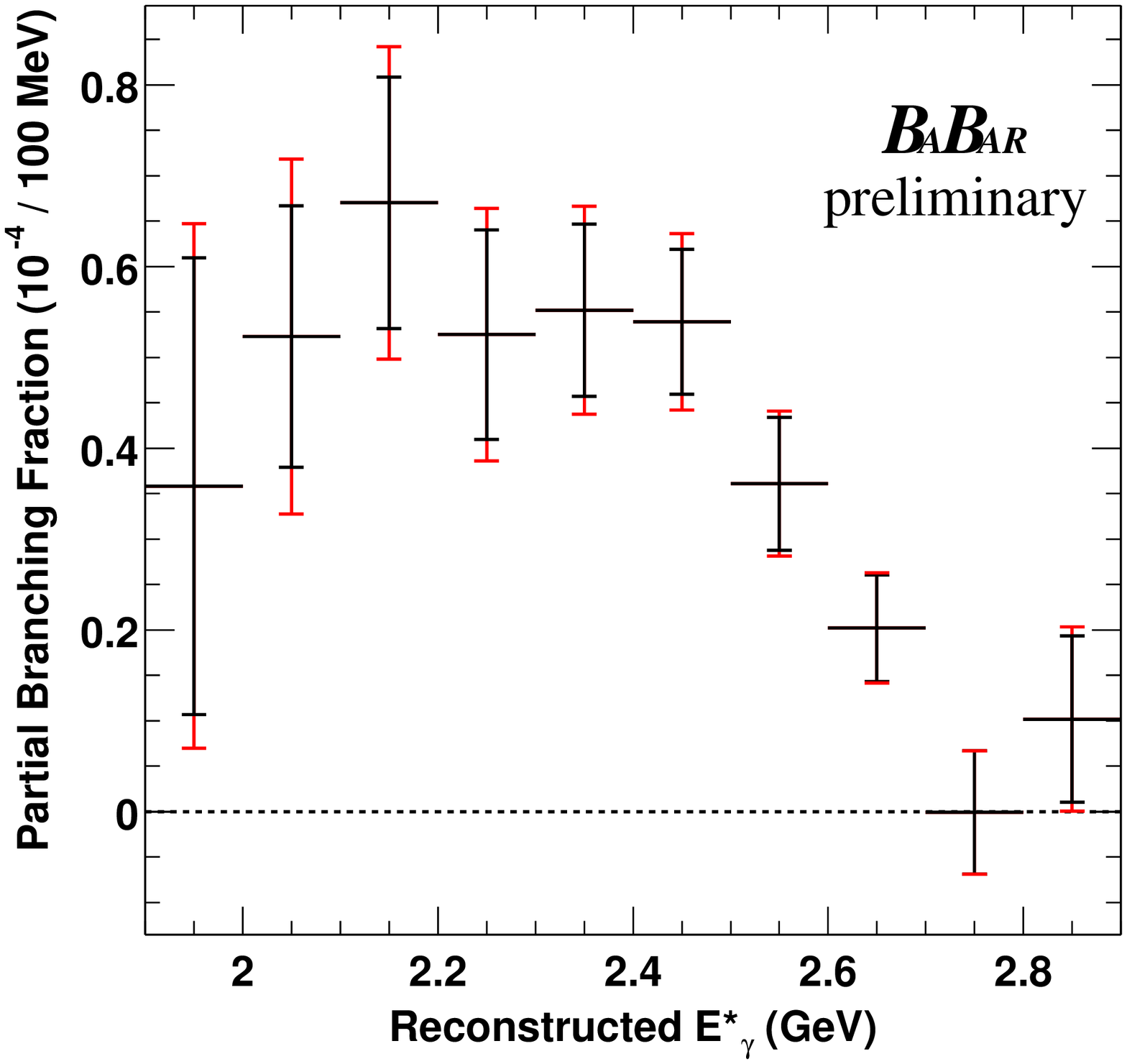}
            \hskip4mm \epsfysize 1.45truein
            \epsfbox{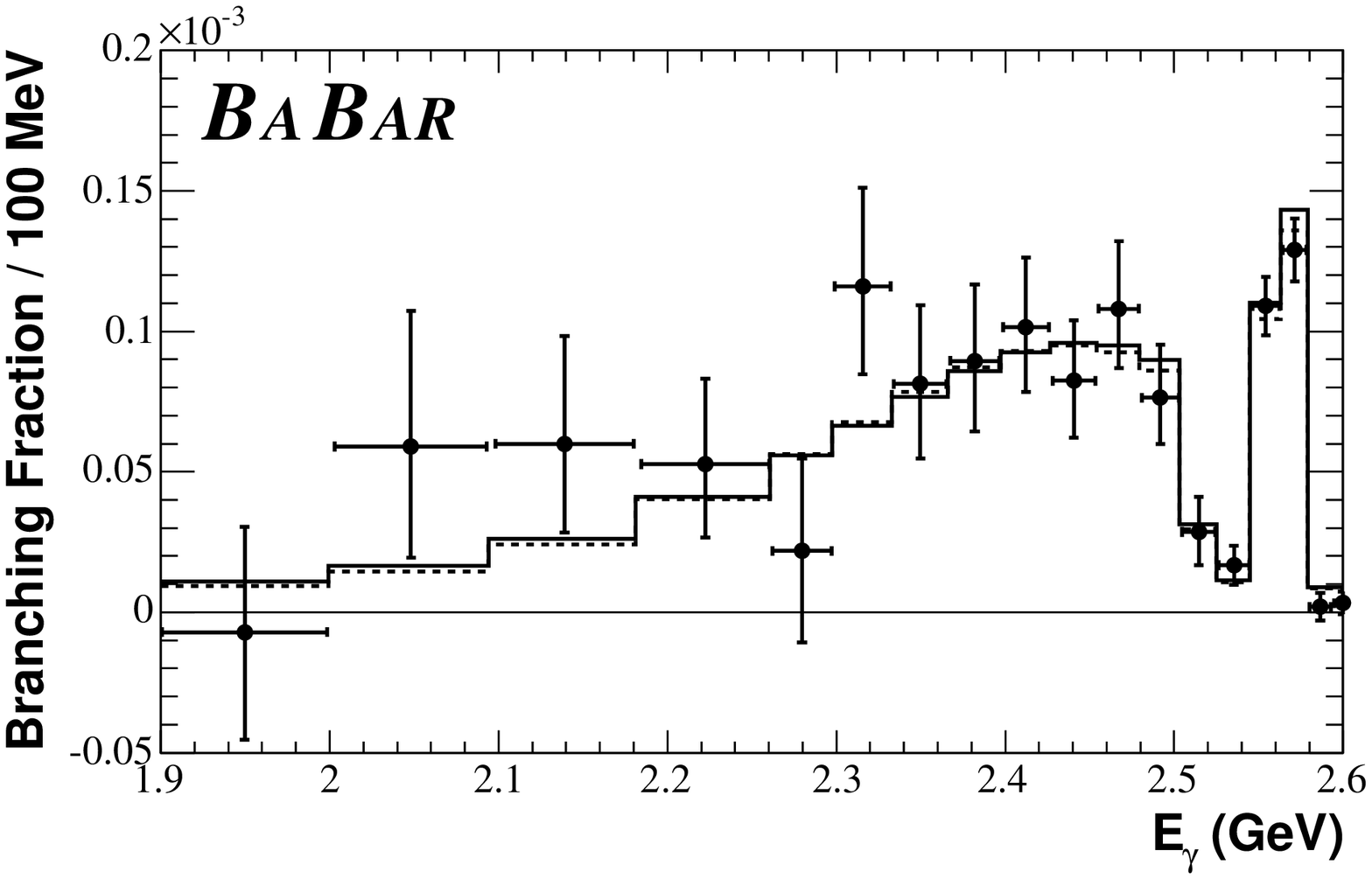}}\vspace{-4mm}
 \caption{\it Photon energy spectrum from fully- (left, in \FourS frame)
              and semi-inclusive $\B\to X_s\g$ analyses (right, in \B
              frame, with theory spectra overlaid).\label{bsg}}
\end{figure}

\begin{figure}[htbp]\vspace{-2mm}
\centerline{\epsfxsize 2.3truein
            \epsfbox{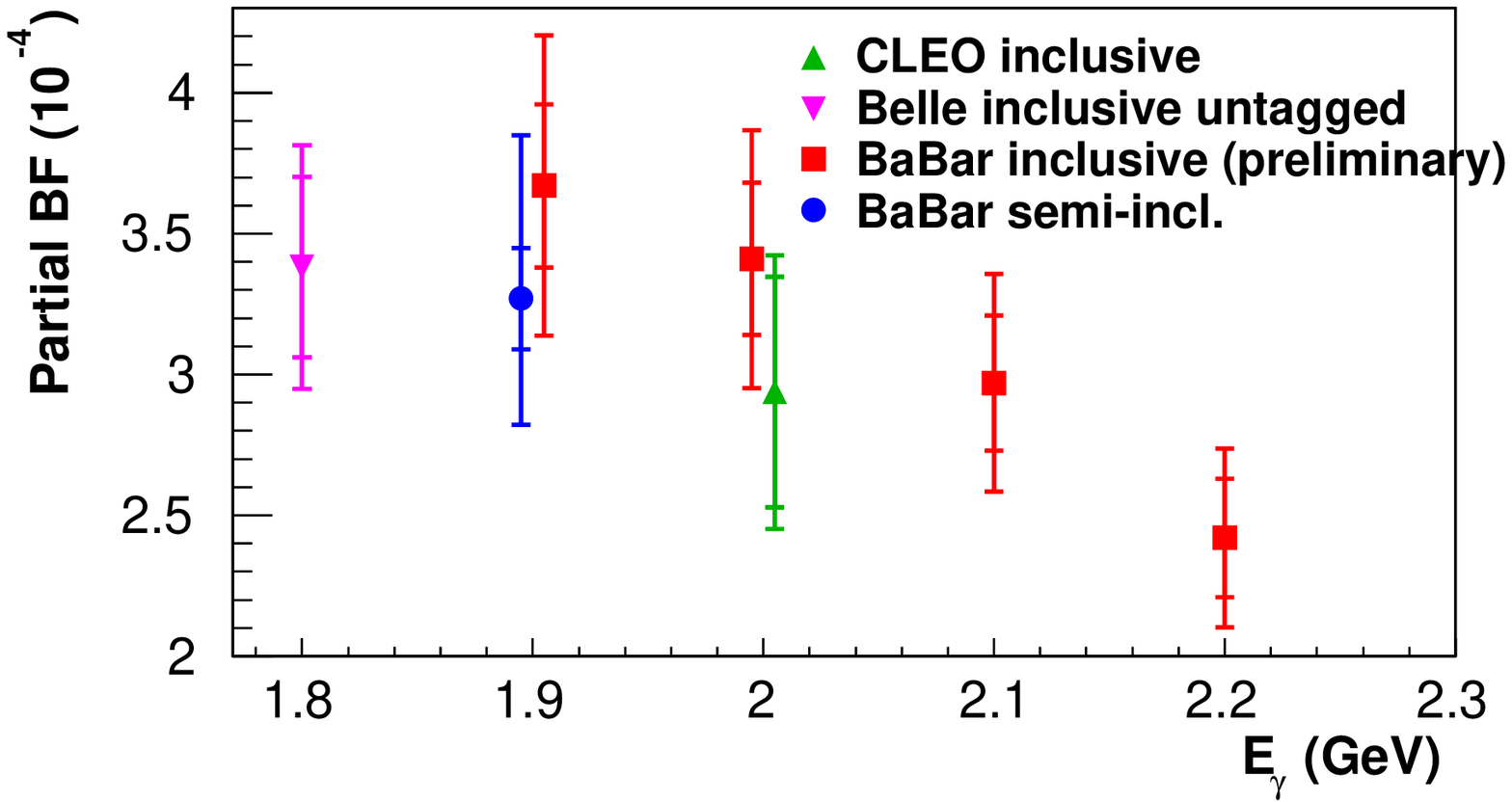}\hskip-2mm
            \epsfxsize 2.5truein
            \epsfbox{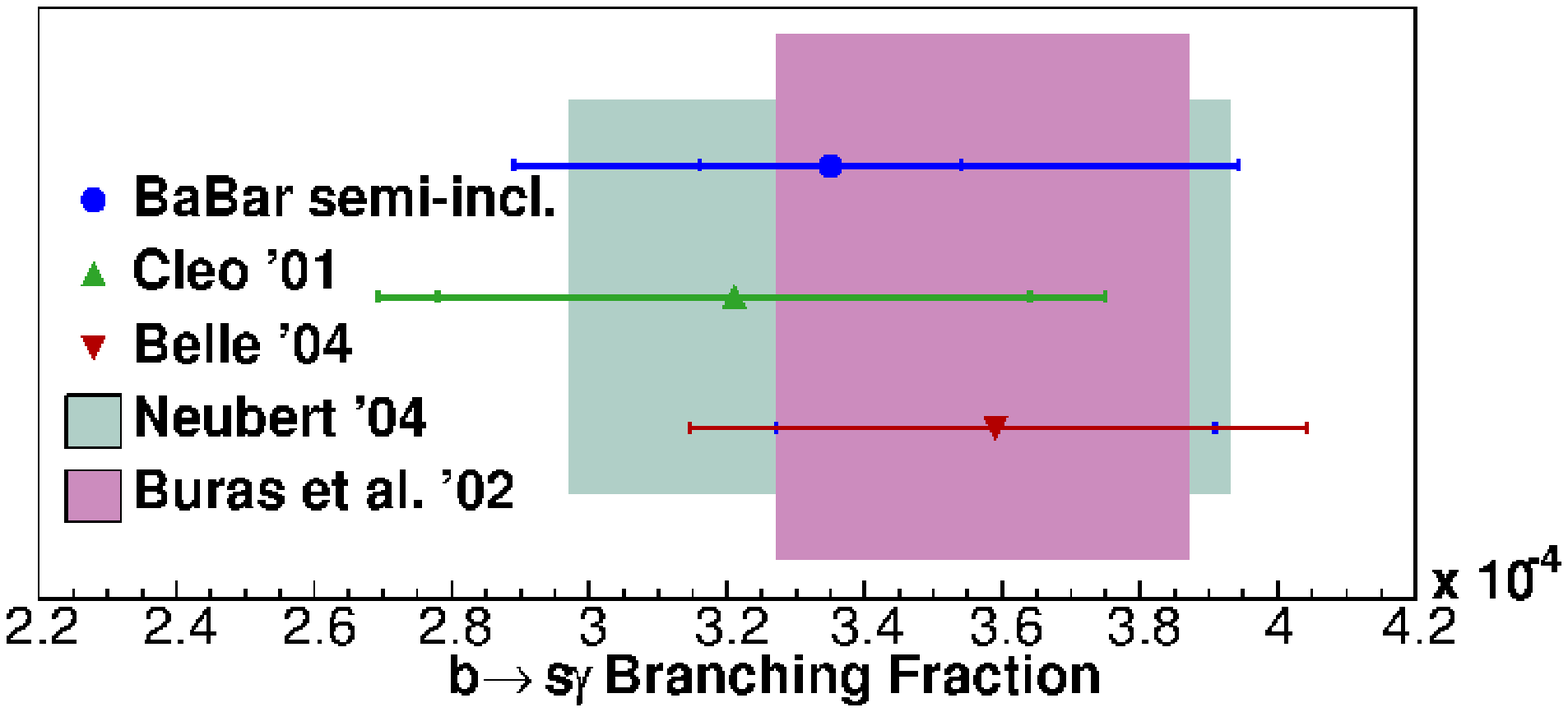}}\vspace{-4mm}
 \caption{\it Partial BFs versus lower cut in $E_\g$ (left) and overall
              BF measurements (right) of $\B\to X_s\g$ for
              $E_\g>1.6\gev$.
    \label{BF} }
\end{figure}

The $E_\g$ spectra from the two $\B\to\X_s\g$ analyses are shown in
Fig.\ref{bsg}.  The~$\Kstar\g$ peak, prominent at high $E_\g$ for the
semi-inclusive analysis, is not visible for the inclusive analysis due
to resolution constraints.  Fig.\ref{BF} left plots the fully-inclusive
partial BFs against the value of the lower cut in $E_\g$.  The~overall
semi-inclusive BF, when extrapolated to $E_\g>1.6\gev$, agrees with the
SM prediction and with the results from other experiments (Fig.\ref{BF}
right).\cite{0507001,0508004}

\begin{figure}[htbp]\vspace{-1mm}
\centerline{
            \epsfysize 1.4truein
            \epsfbox{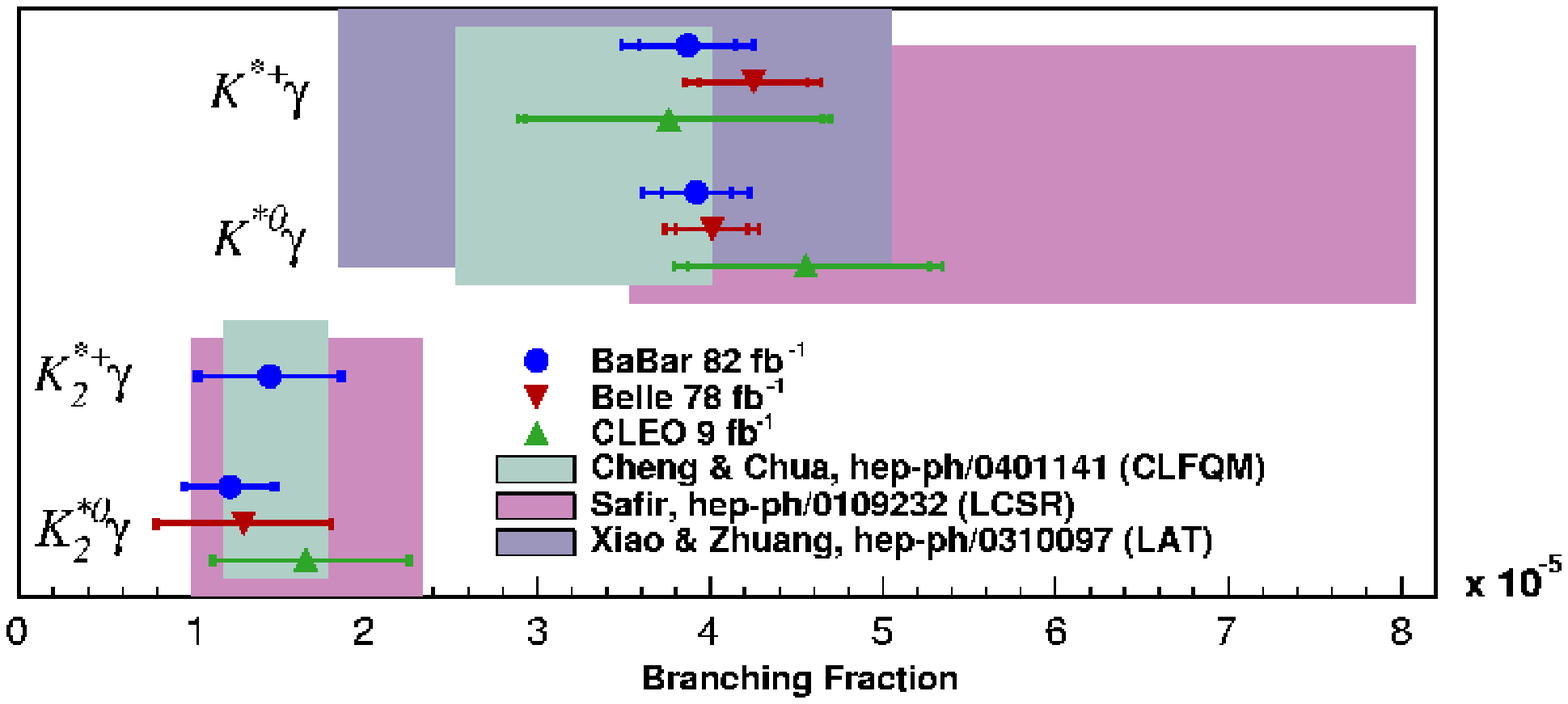}}\vspace{-4mm}
 \caption{\it Branching fractions of $\B\to\Kstar(892)\g$ and
              $\B\to\Kstar_2(1430)\g$.
    \label{kstg} }
\end{figure}

Non-perturbative hadronic effects complicate the theoretical
calculations of exclusive decays like $\B\to\Kstar(892)\g$ and
$\B\to\Kstar_2(1430)\g$, so that the measurements are currently more
accurate than the predictions.  A~summary of the results is shown in
Fig.\ref{kstg}.\cite{0112006,0091105}

\section{$\B\to X_sll$, $\B\to K^{(*)}ll$ and $\B\to(\rho,\omega)\g$}

\begin{figure}[htbp]\vspace{-0mm}
\centerline{\epsfysize 1.2truein
       \epsfbox{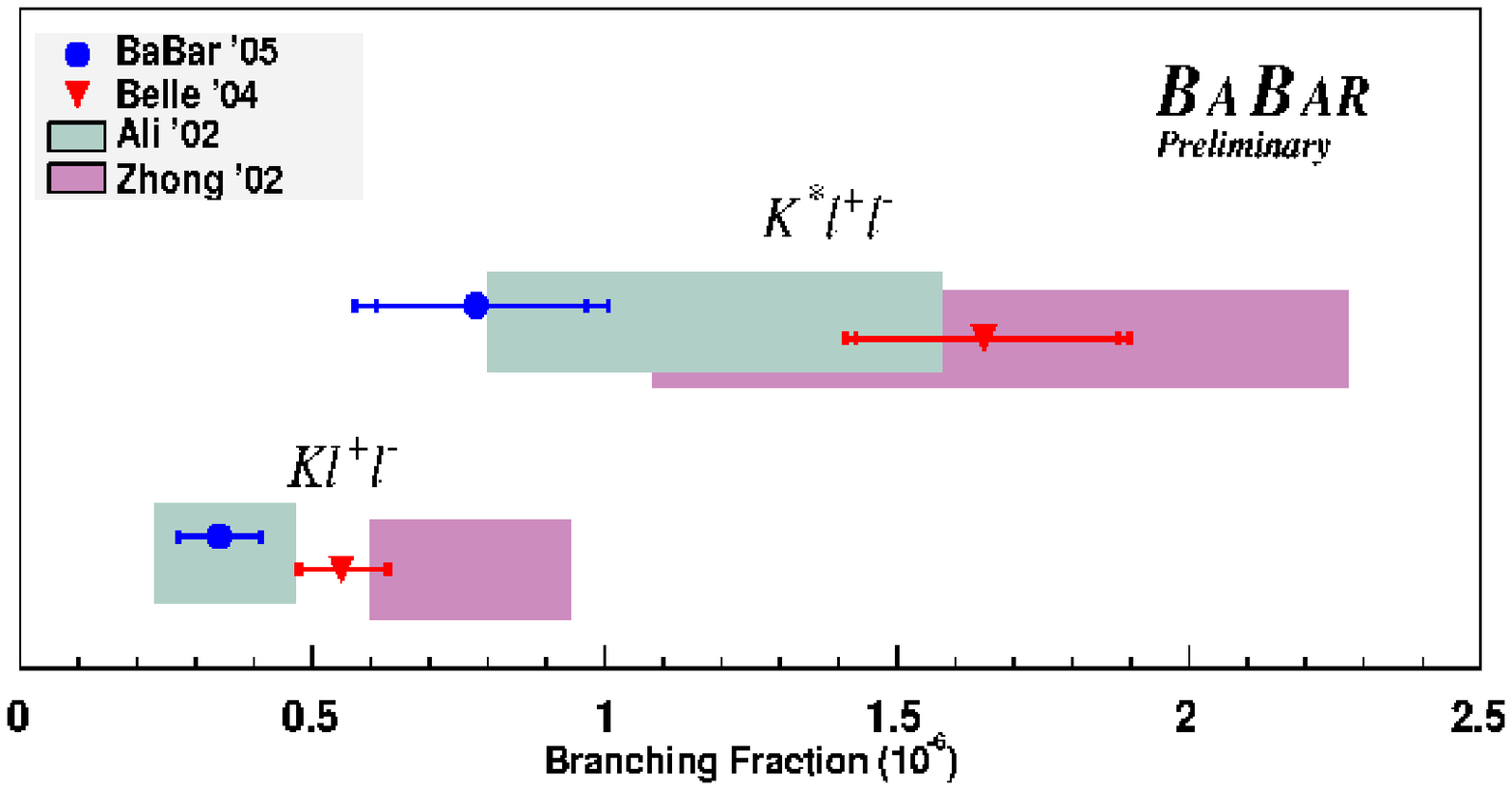}
            \hskip-3mm \epsfysize 1.2truein
            \epsfbox{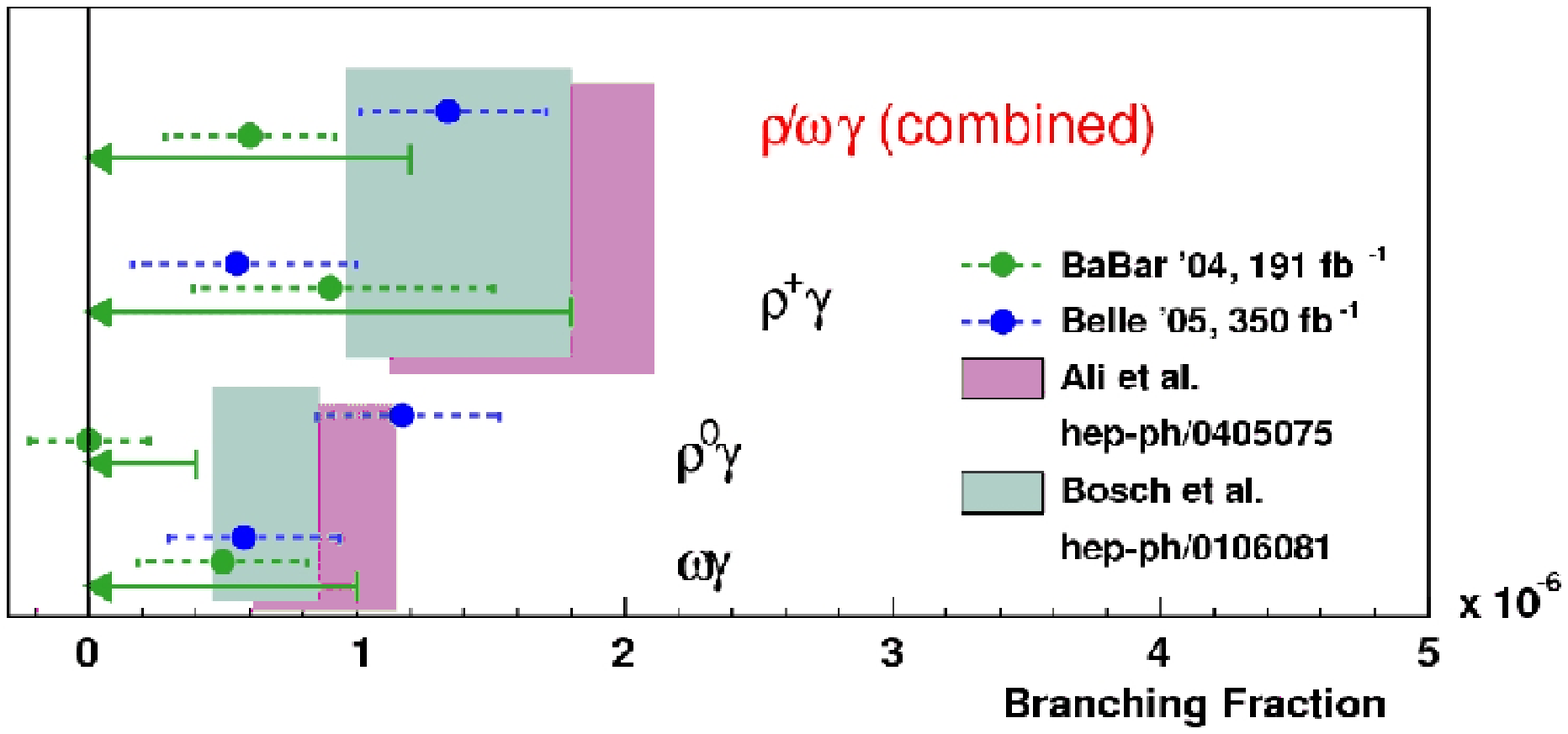}}\vspace{-4mm}
 \caption{\it BF measurements and SM predictions for $K^{(*)}ll$ (left)
              and $\B\to(\rho,\omega)\g$ decays (right).
    \label{LLrhogam} }
\end{figure}

The decay $b\to sll$ has been measured semi-inclusively ($\B\to X_sll$)
on $89$M \BB pairs, and exclusively ($\B\to\kaon^{(*)}ll$) on $229$M \BB
pairs.  The~former measurement is again based on a sum of exclusive
states, with about half of the total rate missing, and its
BF\cite{0081802} of $(5.6\pm1.5\pm0.6\pm1.1)\times10^{-6}$ for
$m_{ll}>0.2\gevcc$ agrees well with the SM prediction.  The exclusive
decay results are shown in Fig.\ref{LLrhogam} left.\cite{0507005}

The decay $b\to\d\g$ has been studied in 221M \BB pairs by searching for
$\B\to(\rho,\omega)\g$.  These decays go primarily through penguin
diagrams, but also through \W-exchange or \W-annihilation.  The
background originates mainly from \qqbar (=$udsc$) events.  The~BF
results are summarized in Fig.\ref{LLrhogam} right.\cite{0011801}

\section{ $\Bzb\to\Dstarz\g$ and $\Bz\to\phi\g$}

\begin{figure}[htbp]\vspace{-4.5mm}
\centerline{\epsfysize 1.37truein \epsfbox{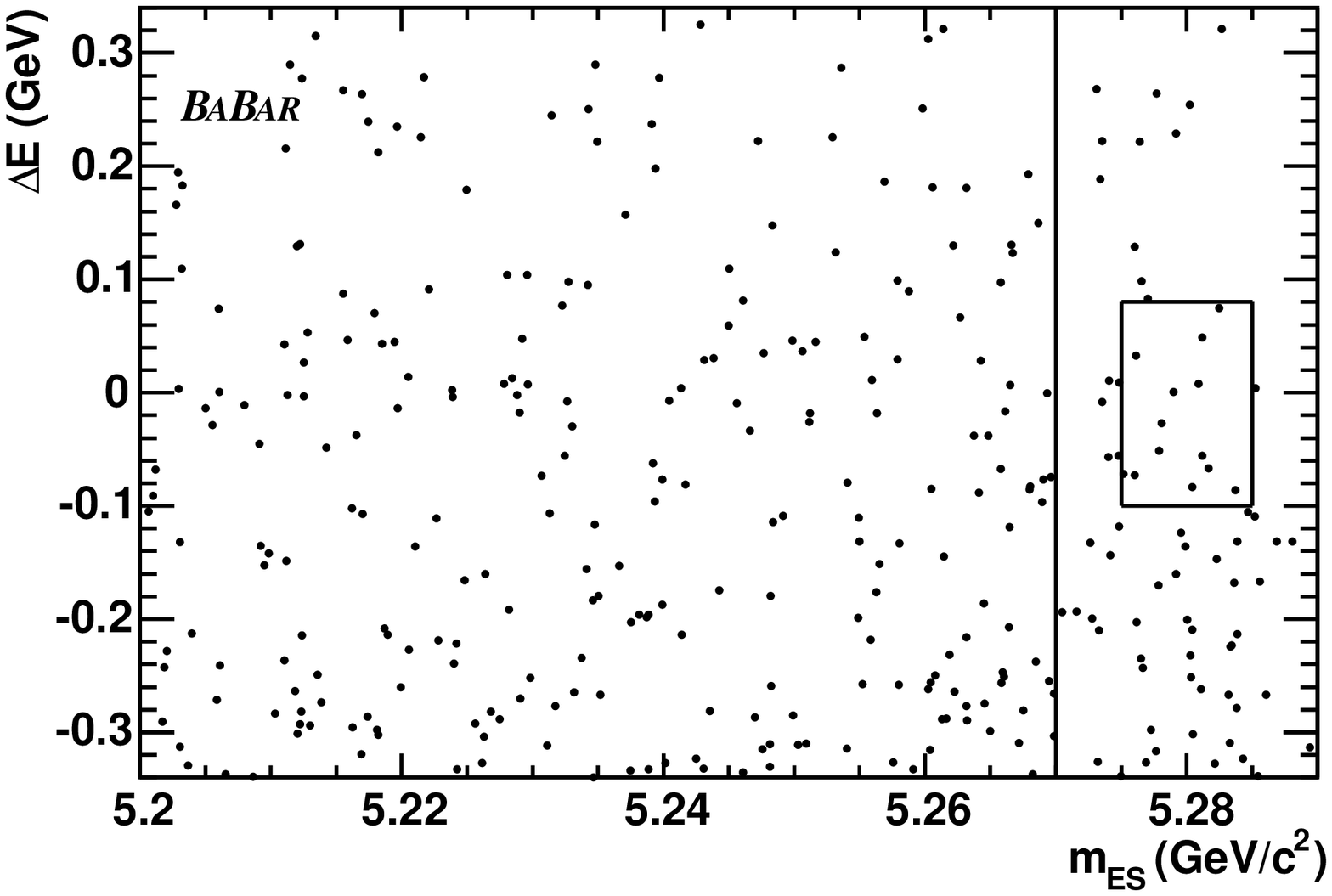}
            \epsfysize 1.43truein
            \epsfbox{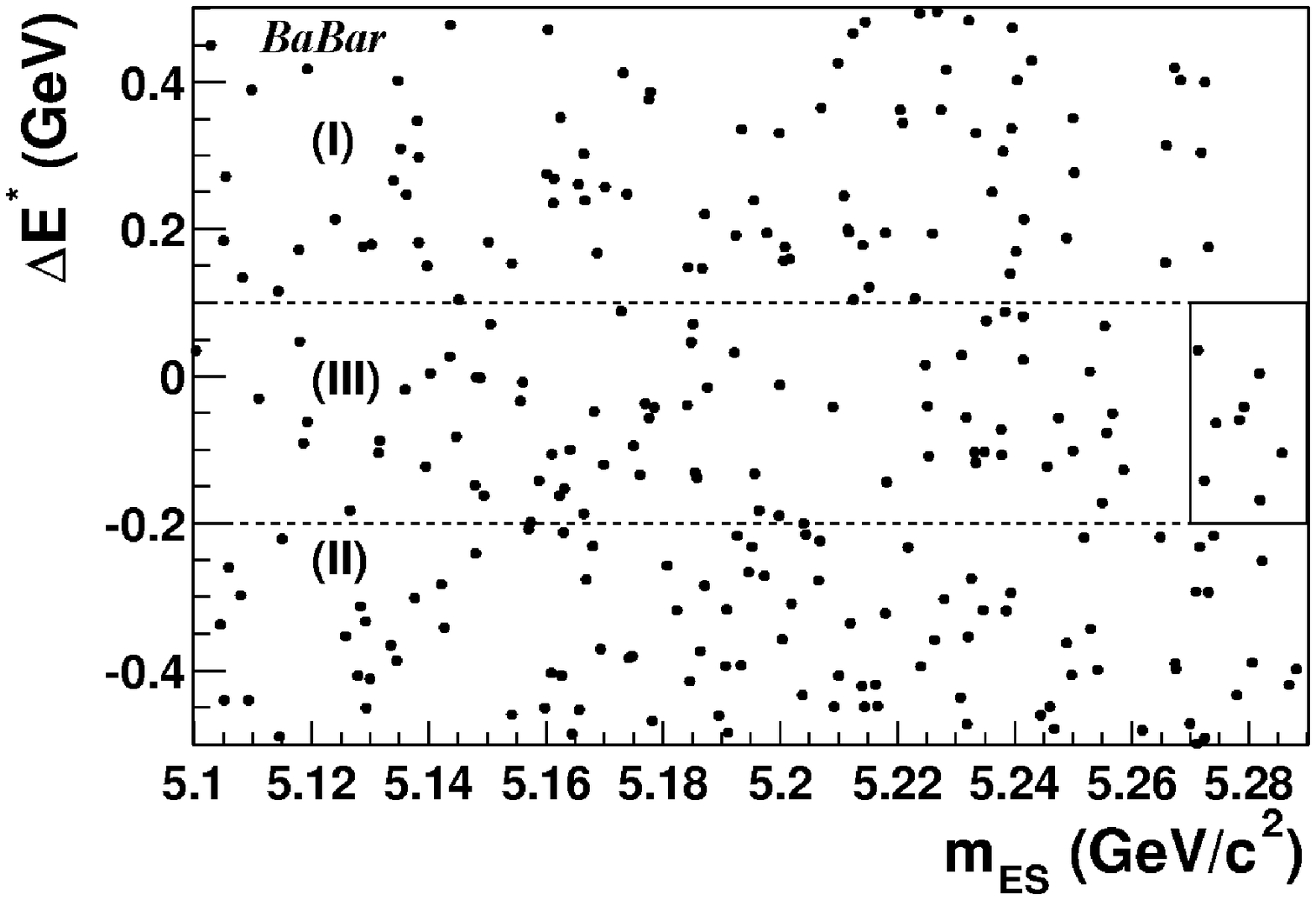}}\vspace{-5mm}
 \caption{\it \mes-\DeltaE plane of real data for $\Bzb\to\Dstarz\g$
              (left) and $\Bz\to\phi\g$ (right).  In~both plots the
              signal box is indicated on the right side.
    \label{Dstar0gphigam} } \vspace{-2mm}
\end{figure}

The~$\Bzb\to\Dstarz\g$ decay with SM predictions around $10^{-6}$ is
dominated by $W$-exchange.  The~final \B candidates from $88$M \BB pairs
are described by $\mes = \sqrt{ E^{*2}_{\rm beam} - p^{*2}_B }$ and
$\DeltaE^* = E^*_B - E^*_{\rm beam} $, with $E^*_{\rm beam}$ being the
center-of-mass (CM) beam energy, and $E^*_B$ and $p^{*2}_B$ the \B
candidate's CM energy and momentum.  Background, mainly from \BB decays,
is estimated to be $9.4\pm1.7$ events in the \mes-\DeltaE signal box.
Thirteen observed data events (Fig.\ref{Dstar0gphigam} left) lead to a
BF upper limit of $2.5\times10^{-5}$ at 90\% confidence level
(CL).\cite{0506070}

The experimental signature of the $\Bz\to\phi\g$ decay is clean, but the
SM prediction of the BF is very low with $3.6\times10^{-12}$.
Candidates are selected from $124$M \BB pairs.  In~the signal region, a
$\qqbar$ (\BB) background of $6.0\pm1.0$ ($<$0.1) events is expected.
Eight events observed in data (Fig.\ref{Dstar0gphigam} right) result in
a BF upper limit of $8.5\times10^{-7}$ at 90\% CL.\cite{0501038}

The~author thanks the \BaBar collaboration, the SLAC accelerator group
and all contributing computing organizations.  He~was supported
by~U.S. \hbox{Department~of} Energy grant DE-FG05-91ER40622.

\end{document}

%% file: slacpub-cover.tex
  \setlength{\textwidth}{16.5cm}

\def\babar   {\mbox{\slshape B\kern-0.1em{\scriptsize A}\kern-0.1em
              B\kern-0.1em{\scriptsize A\kern-0.2em R}}\xspace}

\normalsize

\begin{flushright}
  hep-ex/0607041     \\
  SLAC-PUB-11856     \\
  \babar-TALK-05/113 \\
  July 2006          \\ 
\end{flushright}

\par\vskip 2.0cm

% Title of the paper
\begin{center}
\normalsize \bf 
Rare Decays and Search for New Physics with \babar
\end{center}
\bigskip

\begin{center}
\normalsize Johannes M. Bauer \\ [1mm] for the \babar Collaboration\\ \mbox{ }\\
%\today
\end{center}
\bigskip \bigskip

% Abstract
\begin{center}
\normalsize \bf Abstract
\end{center}
    Rare \B decays permit stringent tests of the Standard Model and
    allow searches for new physics.  Several rare radiative-decay
    studies of the \B meson from the \babar collaboration are described.
    So~far no~sign for new physics was discovered.
\vfill
\begin{center}

\bigskip\bigskip
Submitted to the Conference Proceedings of the Fourth International
Conference on Frontier Science --- New Frontiers in Subnuclear Physics,
September 12--17, 2005, Milan, Italy

\end{center}

\vspace{1.0cm}
\begin{center}
{\em Stanford Linear Accelerator Center, Stanford University, 
Stanford, CA 94309} \\ \vspace{0.1cm}\hrule\vspace{0.1cm}
Work supported in part by Department of Energy contract DE-AC03-76SF00515.
\end{center}

\newpage

%% file: paper.bbl
\vspace{-1mm}

\begin{thebibliography}{99}

\def\refup{\vspace{-3.35mm}}

\refup

\bibitem{NandoGagan}Presentations by Fernando Ferroni and Gagan Mohanty
    this conference.\refup

%%CITATION = HEP-EX 0507001;%%
\bibitem{0507001}\refBaBar Collaboration, B. Aubert \etal, hep-ex/0507001 (2005).\refup

%%CITATION = HEP-EX 0508004;%%
\bibitem{0508004}\refBaBar Collaboration, B. Aubert \etal, \jprd{72}, 052004 (2005).\refup

%%CITATION = HEP-EX 0407003;%%
\bibitem{0112006}\refBaBar Collaboration, B. Aubert \etal, \jprd{70}, 112006 (2004).\refup

%%CITATION = HEP-EX 0409035;%%
\bibitem{0091105}\refBaBar Collaboration, B. Aubert \etal, \jprd{70}, 091105 (2004).\refup

%%CITATION = HEP-EX 0404006;%%
\bibitem{0081802}\refBaBar Collaboration, B. Aubert \etal, \jprl{93}, 081802 (2004).\refup

%%CITATION = HEP-EX 0507005;%%
\bibitem{0507005}\refBaBar Collaboration, B. Aubert \etal, hep-ex/0507005 (2004).\refup

%%CITATION = HEP-EX 0408034;%%
\bibitem{0011801}\refBaBar Collaboration, B. Aubert \etal, \jprl{94}, 011801 (2005).\refup

%%CITATION = HEP-EX 0506070;%%
\bibitem{0506070}\refBaBar Collaboration, B. Aubert \etal, \jprd{72}, 051106 (2005).\refup

%%CITATION = HEP-EX 0501038;%%
\bibitem{0501038}\refBaBar Collaboration, B. Aubert \etal, \jprd{72}, 091103 (2005).


\end{thebibliography}
